\documentclass[11pt,a4paper]{article}

\usepackage{jheppub}
\usepackage{amsfonts}
\usepackage{amssymb}
\usepackage{comment}
\usepackage{epsfig}
\usepackage{graphicx}
\usepackage{amsmath}
\usepackage{tabu}
\usepackage{subfig}
\usepackage{float}
\usepackage{epstopdf}

\newcommand{\bea}{\begin{eqnarray}}
\newcommand{\eea}{\end{eqnarray}}
\newcommand{\bi}{\begin{itemize}}
\newcommand{\ei}{\end{itemize}}
\newcommand{\ben}{\begin{enumerate}}
\newcommand{\een}{\end{enumerate}}
\newcommand{\be}{\begin{equation}}
\newcommand{\ee}{\end{equation}}
\newcommand{\ba}{\begin{align}}
\newcommand{\ea}{\end{align}}
\newcommand{\comments}[1]{}

\def\SM{{\scriptscriptstyle \rm SM}}

\def\KK{{\scriptscriptstyle \rm KK}}
\def\DM{{\scriptscriptstyle \rm DM}}
\def\dS{{\scriptscriptstyle \rm dS}}

\def\H{{\scriptscriptstyle \rm H}}
\def\B{{\scriptscriptstyle \rm B}}

\newcommand\vo{{\mathcal{V}}}

\newcommand{\mc}{\mathcal}
\newcommand{\gsim}{\gtrsim}
\newcommand{\lsim}{\lesssim}

\newcommand{\beqa}{\begin{eqnarray}}
\newcommand{\eeqa}{\end{eqnarray}}

\title{Affleck-Dine Baryogenesis in Type IIB String Models}

\author[1]{Rouzbeh Allahverdi,}
\author[2,3,4]{Michele Cicoli,}
\author[2,3]{Francesco Muia}

\affiliation[1]{Department of Physics and Astronomy,\\ University of New Mexico,\\ Albuquerque, NM 87131, USA}
\affiliation[2]{Dipartimento di Fisica e Astronomia, Universit\`a di Bologna, \\ via Irnerio 46, 40126 Bologna, Italy}
\affiliation[3]{INFN, Sezione di Bologna, via Irnerio 46, 40126 Bologna, Italy}
\affiliation[4]{ICTP, Strada Costiera 11, Trieste 34014, Italy}

\emailAdd{rouzbeh@unm.edu}
\emailAdd{mcicoli@ictp.it}
\emailAdd{muia@bo.infn.it}

\abstract{We present a viable string embedding of Affleck-Dine baryogenesis in type IIB sequestered models where the late-time decay of the lightest modulus reheats the universe to relatively low temperatures. We show that if inflation is driven by a blow-up K\"ahler modulus, the Affleck-Dine field can become tachyonic during inflation if the K\"ahler metric for matter fields has an appropriate inflaton-dependent contribution. We find that the Affleck-Dine mechanism can generate the observed baryon asymmetry for natural values of the underlying parameters which lead also to successful inflation and low-energy gaugino masses in a split supersymmetry scenario. The reheating temperature from the lightest modulus decay is high enough to allow thermal Higgsino-like dark matter.}


\begin{document}

\maketitle

\section{Introduction}
\label{sec:introduction}

The origin of the matter-antimatter asymmetry of the universe still remains a mystery to be unraveled~\cite{BAU}. Most current approaches are based on mechanisms that rely on relatively high reheating temperatures $T_{\rm rh}$. For example, thermal leptogenesis~\cite{lepto} requires $T_{\rm rh}\gtrsim 10^{9}$-$10^{10}$ GeV while electroweak baryogenesis~\cite{EW} demands $T_{\rm rh} \gtrsim 1$ TeV. 

However, supersymmetric (SUSY) extensions of the Standard Model (SM) and their string theory embeddings typically have moduli fields that alter the standard thermal history of the universe~\cite{ksw}. The moduli, due to their gravitational coupling to matter, are long-lived and tend to dominate the energy density of the universe before decaying. The late-time decay of the moduli typically gives rise to reheating temperatures well below the electroweak scale, particularly in models with low-energy SUSY. The moduli decay also releases a huge amount of entropy that dilutes any pre-existing relic abundance, thereby necessitating the production of dark matter (DM) and the generation of the baryon asymmetry of the universe (BAU) at relatively low temperatures. 

An interesting class of string compactifications is type IIB sequestered string models with D3-branes at singularities \cite{SingD3s}. These models have been explicitly embedded in globally consistent Calabi-Yau (CY) compactifications with de Sitter (dS) closed string moduli stabilisation \cite{CYembedding}. Moreover, they can yield low-energy SUSY without introducing any cosmological moduli problem \cite{CMP} or gravitino overproduction problem \cite{gravProbl}, and provide a promising framework for building inflationary models in agreement with Planck data where the inflaton is a K\"ahler modulus \cite{InflReview}. 

In the context of the Large Volume Scenario (LVS) \cite{LVS}, all sequestered models share a universal feature: the overall volume mode is the lightest modulus $\chi$. Its mass is suppressed with respect to the gravitino mass: $m_\chi \sim m_{3/2}\,\sqrt{\epsilon}$, where $\epsilon \sim m_{3/2}/M_p \ll 1$. On the other hand gaugino masses scale as $M\sim m_{3/2}\,\epsilon$ while scalar masses can behave as either $m_0 \sim M$ or $m_0 \sim m_\chi$ \cite{Blumenhagen:2009gk,Aparicio:2014wxa} depending on the exact form of the K\"ahler metric for matter fields and the mechanism responsible for achieving a dS vacuum. Thus sequestered scenarios can give rise to both MSSM-like and split SUSY-like models. TeV-scale gaugino masses can be obtained 
for $m_{3/2}\sim 10^{10}$-$10^{11}$ GeV and $m_\chi \sim 10^{6}$-$10^{7}$ GeV. The decay of $\chi$ typically gives rise to a reheating temperature of order $T_{\rm rh}\sim m_\chi \sqrt{m_\chi/M_p} \sim {\cal O}(10)$ GeV. 

A promising mechanism for generating the observed value of BAU in models with reheating temperatures below the EW scale is Affleck-Dine (AD) baryogenesis~\cite{Affleck:1984fy}. This scenario utilizes a SUSY D-flat direction that carries a non-zero baryon number, the so-called AD field. If the AD field develops a sufficiently large displacement from its late-time minimum during inflation, its post-inflationary dynamics can generate a baryon asymmetry that can survive the entropy release during the final stage of reheating. However, a successful embedding of AD baryogenesis in supergravity models is non-trivial as supergravity corrections can ruin the flatness of the potential for the AD field, thereby preventing it from acquiring a large vacuum expectation value (VEV) during inflation.         

In this paper, we show that BAU can be successfully generated in sequestered models via AD mechanism. In particular, we shall outline how to construct a model where one can follow the whole cosmological evolution of the universe from inflation to the final stage of reheating by decay of the lightest modulus which successfully generates the observed BAU along with the correct DM relic abundance. The highlights of the scenario are as follows:
\bi
\item Small-field inflation takes place in the closed string sector where a blow-up modulus $\sigma$ drives the exponential expansion of the universe 
in agreement with Planck data \cite{Conlon:2005jm}. Generating density perturbations with the correct amplitude raises all the mass scales mentioned above by about two orders of magnitude, which results in gaugino masses in the range $M \sim 10^{4}$-$10^{5}$ GeV. 

\item If the K\"ahler metric for matter fields has an appropriate dependence on $\sigma$, in split SUSY-like models squarks and sleptons can become tachyonic during inflation. The AD field $\phi$ can then develop a sufficiently large non-zero VEV during inflation. Moreover also the volume mode $\chi$ is shifted from its late-time minimum during inflation. 

\item At the end of inflation, the inflaton becomes very heavy since $m_\sigma \sim m_{3/2}$ and its decay leads to an initial stage of reheating with a relatively high temperature. When the Hubble constant $H$ drops to $m_\phi \sim m_\chi$, both $\phi$ and $\chi$ start oscillating around their late-time minima. The SUSY breaking A-terms induce a rotational motion of the AD field and the generation of baryon asymmetry that gets transferred to quarks when $\phi$ decays.

\item Given that $\chi$ is only gravitationally coupled, it decays after $\phi$ and dilutes any previously produced relic abundance.
The final reheating temperature is $T_{\rm rh}\sim 10^{2}$-$10^{4}$ GeV, which is high enough to allow thermal Higgsino-like DM with a mass around $1$ TeV. The observed BAU can be obtained for natural initial displacements of the AD field $\phi$ of order $0.1\,M_p$.
\ei

We would like to emphasize that the sequestered model in this paper represents an explicit example where one can check in detail the viability of the main assumption underlying the AD mechanism for baryogenesis, i.e. the dynamical generation of a tachyonic mass during inflation. In fact, this class of models show an interesting interplay between inflation, SUSY breaking, soft terms, baryogenesis and DM within the same string compactification. 

The paper is organised as follows. In Sec. \ref{sec:ADreview}, we briefly review AD baryogenesis and the present constraints for its embedding in a supergravity framework. In Sec. \ref{SecSeqLVS}, we give an overview of all the main features of sequestered LVS models like the structure of the 4D effective field theory, moduli stabilisation, SUSY breaking and the spectrum of soft terms. In Sec. \ref{sec:dynamics}, we describe the full cosmological evolution in our scenario including inflation, reheating and generation of BAU via the AD mechanism. Finally, we present and discuss our numerical results in Sec. \ref{SecResults} before concluding in Sec. \ref{Concl}.

\section{Review of Affleck-Dine baryogenesis}
\label{sec:ADreview}

\subsection{Basic mechanism}

The field space of SUSY extensions of the SM contains many directions along which the D-term contributions to the scalar
potential identically vanish. These D-flat directions are parameterised by gauge invariant monomials of the chiral superfields~\cite{Dine:1995kz,gkm}. 
They are lifted by SUSY breaking terms and superpotential terms of the form $\lambda_n {\Phi}^n/n \Lambda^{n-3}$. Here $n \geq 3$, $\Lambda$ is the scale of new physics (typically Planck or string scale), and $\Phi$ is the superfield comprising the flat direction $\phi$ and its fermionic partner. 

In general, both the low-energy SUSY breaking and SUSY breaking by the non-zero energy density of the universe contribute to the lifting of flat directions. As a result, the scalar potential along $\phi$ can be written as~\cite{Dine95}:
\be
V(\phi) = \left(m^2_\phi + c_\H H^2 \right) {\vert \phi \vert}^2 +
\vert \lambda_n \vert^2 {{\vert \phi \vert}^{2(n-1)} \over \Lambda^{2(n-3)}} + \left[(A_n + a_n H) {\lambda_n \phi^n \over n \Lambda^{n-3}} + {\rm h.c.} \right]\, ,
\label{pot}
\ee
where $H$ is the Hubble expansion rate. The three terms on the right-hand side of (\ref{pot}) represent, respectively, the sum of the low energy and Hubble induced soft mass terms, the contribution of superpotential terms and the sum of the low-energy and Hubble induced $A$-terms. 

The role of $\phi$ in the early universe\footnote{SUSY flat directions and their cosmological consequences are discussed in detail  in~\cite{flat}.} crucially depends on the size and sign of $c_\H$: 
\vskip 2mm
\noindent
{\bf (1)} $c_\H \gsim 1$. In this case, $\phi$ has a mass that exceeds the Hubble expansion rate during inflation $H_{\rm inf}$. Then, due to fast rolling,
the field settles down to the origin of its potential during inflation, which bears no interesting consequences. 
\vskip 2mm
\noindent
{\bf (2)} $0 < c_\H \ll 1$. In this case $\phi$ makes a quantum jump of length $H_{\rm inf}/2\pi$ within
each Hubble time. These jumps superpose in random walk fashion, resulting in the following maximum displacement for $\phi$ during inflation~\cite{Linde}:
\be 
\phi_{\rm max} = \sqrt{\frac{3}{8 \pi^2}} \frac{H^2_{\rm inf}}{m_\phi}\,.
\label{rms}
\ee
\vskip 2mm
\noindent
{\bf (3)} $c_\H < 0$.\footnote{A negative $c_\H$ can arise at the tree-level~\cite{Dine95}, or from radiative corrections~\cite{gmo,adm}.} In this case $\phi$ becomes tachyonic during inflation. Hence it is driven away from the origin and quickly settles at the minimum of the potential that has a large VEV: 
\be \label{tach}
\phi_{\rm inf} = \left({H_{\rm inf} \Lambda^{n-3} \over \sqrt{n-1} c_\H \lambda_n} \right)^{1/(n-2)}.
\ee
After inflation $\phi$ follows an instantaneous value that slowly varies due to the expansion of the universe. Once $H \sim m_\phi$, low energy SUSY breaking terms take over the potential, see (\ref{pot}), and $\phi$ starts oscillating about the origin with an initial amplitude:
\be \label{initial}
\phi_0 ~ \sim ~ \left({m_\phi \Lambda^{n-3} \over \sqrt{n-1} \lambda_n}\right)^{1/(n-2)}\, .
\ee
The $\phi$ field also feels a torque due to the $A$-term in (\ref{pot}) that results in rotational motion~\cite{Dine95}. If $\phi$ has a non-zero baryon number $\beta$, rotation will result in a baryon asymmetry~\cite{Affleck:1984fy}. The asymmetry per $\phi$ quanta is given by~\cite{Dine95}:
\be \label{asym}
{n_{\B} \over n_\phi} \sim \beta ~ {\rm sin} (n \theta_i) ~ {|A_n| \over m_\phi}\,,
\ee
where $\theta_i$ is the initial angular displacement of $\phi$ from the minimum of the potential in the angular direction. This displacement could be due to a mismatch between the phases of $A$ and $a$, see (\ref{pot}), or a result of initial conditions along the angular direction (if $|a| \ll 1$). The comoving value of the generated baryon asymmetry remains constant at $H \ll m_\phi$ since the $A$-term is redshifted quickly by Hubble expansion. The asymmetry is transferred to fermions when $\phi$ decays and the final value of BAU $n_\B/s$ (with $s$ being the entropy density) depends on the details of the post-inflationary thermal history of the universe. We will give a detailed discussion on the generation of BAU in our model later on.      

\subsection{Supergravity constraints}
\label{ssec:currentcontraints}

The possibility to realise AD baryogenesis in a 4D $N=1$ supergravity theory has already been analysed in \cite{Casas:1997uk, Dutta:2010sg, Marsh:2011ud, Dutta:2012mw}. The requirement that D-flat directions develop a tachyonic mass during inflation but are non-tachyonic in the present vacuum sets strong constraints on the form of the K\"ahler metric. It has therefore been very difficult to build an explicit supergravity inflationary model where the AD field acquires a large VEV during inflation and relaxes at the origin at later times. 
In order to overcome these difficulties, let us first summarise the common assumptions behind the previous analyses:
\ben
\item The Hubble-induced mass of the AD field $\phi$ during inflation is determined by the F-term of just one field, the inflaton $\sigma$, which dominates the energy density;
\item The gravitino mass is never larger than the Hubble scale during inflation, i.e. $m_{3/2} \lesssim H_{\rm inf}$;
\item The contribution of the F-term of $\sigma$ to the scalar mass of $\phi$ is negligible after the end of inflation in order to avoid tachyons.
\een
Starting from these assumptions, a no-go theorem has been worked out in~\cite{Dutta:2012mw} stating that in the case with $m_{3/2}\ll H_{\rm inf}$ it is impossible to realise both inflation and AD baryogenesis if the holomorphic sectional curvature at a given point of the field space is constant for all choices of planes in the tangent space. Moreover, it has been found~\cite{Casas:1997uk, Dutta:2010sg, Marsh:2011ud} that $\phi$ cannot become tachyonic during inflation if the K\"ahler metric for matter fields $\tilde{K}$ scales exactly as $\tilde{K} = e^{K/3}$ where $K$ is the K\"ahler potential for the geometric moduli. 

In this paper we shall show how to construct an explicit model that can successfully accommodate inflation and AD baryogenesis by relaxing the previous assumptions as follows:
\ben
\item The Hubble-induced mass of $\phi$ during inflation is determined by the F-terms of several moduli and not just the inflaton $\sigma$;
\item The gravitino mass is larger than the Hubble scale during inflation, $m_{3/2} \gg H_{\rm inf}$, while low-energy SUSY is obtained since the visible sector is sequestered from the sources of SUSY breaking;
\item The F-term of $\sigma$ contributes to the mass of $\phi$ also after the end of inflation in a way compatible with $m_\phi^2>0$;
\item We consider cases where the relation $\tilde{K} = e^{K/3}$ is satisfied only at leading order while it is broken by subleading corrections.
\een

\section{Sequestered Large Volume Scenario}
\label{SecSeqLVS}

\subsection{Setup of the compactification}
\label{sec:setup}

String compactifications always come with a number of energy scales which are dynamically determined by the theory in terms of a the single fundamental scale: the string length $\ell_s = M_s^{-1}$, where $M_s$ is the string scale. We denote the volume of the compact space $X$ by $\vo$ (measured in string units in 4D Einstein frame). The low-energy supergravity approximation holds when $\vo \gg 1$, producing the following hierarchy of scales:
\be
M_\KK \simeq \frac{M_p}{\sqrt{4 \pi} \vo^{2/3}} \quad \ll \quad M_s = \frac{g_s^{1/4} M_p}{\sqrt{4 \pi \vo}} \quad \ll \quad M_p\,,
\ee
where $M_\KK$ is the Kaluza-Klein scale, while $M_p$ is the reduced Planck mass and $g_s$ is the string coupling constant. We are interested in the type IIB low-energy theory below $M_\KK$ which is characterised by the presence of many moduli: 
\bi
\item The axio-dilaton $S$ whose real part $s$ determines the string coupling $g_s = \langle s \rangle^{-1}$.

\item $h^{1,2}$ complex structure moduli $U_\alpha$ which parameterise the shape of the internal space.

\item $h^{1,1}$ K\"ahler moduli $T_i$ whose real parts $\tau_i = \text{Re} (T_i) = \text{Vol} (D_i)$ give the size of the divisors $D_i$ of the compact space. The imaginary parts are instead axion-like fields $\psi_i = \text{Im} (T_i) = \int_{D_i} C_4$ where $C_4$ is the Ramond-Ramond 4-form.
\ei

We work in the framework of LVS models where the CY volume takes a typical Swiss-cheese form:
\be
\label{eq:volume}
\vo = \tau_b^{3/2} - \sum_{i=1}^{n} \tau_i^{3/2}\,.
\ee
The overall volume $\vo$ is controlled by the size $\tau_b$ of a `big' divisor $D_b$. The internal space $X$ contains also $n$ `holes', the `small' divisors $D_i$, whose size is much smaller than that of $D_b$: $\tau_i \ll \tau_b$. String compactifications that we are going to discuss include O3/O7-planes and the following small cycles \cite{CYembedding, ModStabChir}:
\bi
\item $n$ divisors $D_i$, $i=1,...,n$, supporting non-perturbative effects and allowing moduli stabilisation \`a la LVS. As we shall explain in Sec.~\ref{ssec:KMI}, the modulus governing the size of the $n$-th divisor, $\tau_n = \text{Vol} (D_n)$, will play the r\^ole of the inflaton.

\item A couple of shrinkable divisors $D_a$ and $D_b$ which are exchanged by the orientifold involution. This results in a linear combination $D_+ = D_a + D_b$ which is even under the orientifold involution and a combination $D_- = D_a - D_b$ which is instead odd. $D_+$ is the cycle which hosts the visible sector and its size is denoted by $\tau_\SM = \text{Vol}(D_+)$, while $D_-$ gives rise to an additional K\"ahler modulus $G = \int_{D_-} B_2 + i \int_{D_-} C_2$, where $B_2$ and $C_2$ are respectively the Kalb-Ramond and the Ramond-Ramond 2-forms belonging to the massless spectrum of type IIB string theory. Both $D_+$ and $D_-$ shrink to zero size due to D-terms stabilisation, namely $\tau_\SM, \text{Re}(G) \rightarrow 0$, while the corresponding axions are eaten up by anomalous $U(1)$s \cite{CYembedding}. 
\ei

The low-energy effective field theory is an $N=1$ supergravity which is completely specified in terms of the real K\"ahler potential $K$, the holomorphic superpotential $W$ and the gauge kinetic functions $f_a$ ($a$ refers to different gauge groups). The K\"ahler potential reads (including the leading $\alpha'$ correction proportional to the constant $\hat\xi$ \cite{BBHL}):
\be
\label{eq:kahlerpotential}
K = - 2 \ln \left(\vo + \frac{\hat\xi}{2}\right) - \ln (S + \overline{S}) + \frac{\tau_\SM^2}{\vo} + \frac{b^2}{\vo} + K_{\rm cs}(U) + K_{\rm matter} \,,
\ee
where we defined $b \equiv \text{Re}(G)$ and $\hat\xi = \xi s^{3/2}$. $K_{\rm cs}(U)$ is the tree-level K\"ahler potential for the $U$-moduli and $K_{\rm matter}$ is the K\"ahler matter metric which includes the visible fields $C^\alpha$. It takes the generic form \cite{Blumenhagen:2009gk,Aparicio:2014wxa}:
\be
\label{eq:kahlermattermetric1}
K_{\rm matter} = \tilde{K}_\alpha(U,S,T) C^\alpha \overline{C}^{\overline{\alpha}} + Z(U,S,T) \left(H_u H_d + \text{h.c.}\right) \,.
\ee
As we shall explain in Sec.~\ref{sssec:scalarmasses}, most of the results of the present paper depend on the exact form of $\tilde{K}_\alpha$. Its leading order $\vo$-scaling can be easily inferred by a locality argument: since visible sector fields are localised at a singularity, the physical Yukawa couplings $\hat{Y}_{\alpha \beta \gamma}$ should not depend (at least at leading order) on the overall volume $\vo$ \cite{Conlon:2006tj}. They are related to the holomorphic Yukawas $Y_{\alpha \beta \gamma}(U,S)$ as:
\be
\hat{Y}_{\alpha \beta \gamma} = e^{K/2} \frac{Y_{\alpha \beta \gamma}(U,S)}{\sqrt{\tilde{K}_\alpha \tilde{K}_\beta \tilde{K}_\gamma}} \,.
\ee
Notice that at perturbative level $Y_{\alpha \beta \gamma}(U,S)$ do not depend on the $T$-moduli due to the holomorphicity of $W$ and the perturbative shift-symmetry of the axions $\psi_i$. Requiring that $\hat{Y}_{\alpha \beta \gamma}$ do not depend on $\vo$ implies that:
\be
\tilde{K}_\alpha = f_\alpha(U,S)\, e^{K/3} \,,
\label{eq:uldefinition}
\ee
where $f_\alpha(U,S)$ is a generic flux-dependent function. We distinguish between two limits, along the lines of \cite{Aparicio:2014wxa}:
\ben
\item \textit{Ultra-local limit}:~\eqref{eq:uldefinition} holds at all orders in the $\vo$-expansion.

\item \textit{Local limit}:~\eqref{eq:uldefinition} holds only at leading order in the $\vo$-expansion.
\een
Given that the factor $e^{K/3}$ in (\ref{eq:uldefinition}) can be expanded as (for vanishing VEVs of $\tau_\SM$, $b$ and matter fields):
\be
e^{K/3}  = \frac{e^{K_{\rm cs}/3}}{(2\,s)^{1/3}}\frac{1}{\left(\vo + \frac{\hat\xi}{2}\right)^{2/3}} = 
\frac{e^{K_{\rm cs}/3}}{(2\,s)^{1/3}}\frac{1}{\hat\vo^{2/3}}\left(1+\frac 23 \frac{\tau_n^{3/2}}{\hat\vo} -\frac{\hat\xi}{3\hat\vo}+ \cdots\right)\,,
\ee
where $\hat\vo = \tau_b^{3/2} - \sum_{j=1}^{n-1}\tau_j^{3/2}$, the K\"ahler matter metric $\tilde{K}_\alpha$ can be parameterised as:
\be
\tilde{K}_\alpha = \frac{f_\alpha(U,S)}{\hat\vo^{2/3}}\left(1- 2 c_n \,\frac{\tau_n^{3/2}}{\hat\vo} -c_\xi \frac{\hat\xi}{\hat\vo}+ \cdots\right)\,.
\ee
The ultra-local limit defined above would then correspond to the case $c_\xi = - c_n = 1/3$.

The holomorphic superpotential is instead given by:
\be
\label{eq:superpotential}
W = W_0(U,S) + \sum_{i=1}^n A_i(U,S)\, e^{-a_i T_i} + W_\dS + W_{\rm matter}\,,
\ee
where $W_0(U,S)$ is the flux-dependent tree-level superpotential \cite{Gukov:1999ya}:
\be
W_0(U,S) = \int_X G_3 \wedge \Omega\,,
\ee
where $\Omega$ is the holomorphic 3-form of $X$ while $G_3 = F_3 - S H_3$, with $F_3 = dC_2$ and $H_3 = dB_2$. $A_i(U,S)$, $i=1,...,n$, are $\mc{O}(1)$ coefficients which depend on the $S$ and $U$-moduli.\footnote{From now on we will omit the $U$ and $S$ dependence in both $W_0$ and $A_i$.} For the coefficients $a_i$ we shall take $a_i = 2 \pi/N_i$ with $N_i \in \mathbb{N}$. Finally $W_\dS$ contains information about the dS sector, while $W_{\rm matter}$ depends on the matter fields $C^\alpha$ and looks like:
\be
\label{eq:wmatter}
W_{\rm matter} = \mu(U,S,T) H_u H_d + Y_{\alpha \beta \gamma}(U,S,T) C^\alpha C^\beta C^\gamma + \dots\,,
\ee
where the dots indicate higher order terms in the expansion around the VEV of visible fields $\langle C^\alpha \rangle = 0$. $\mu$ is the $\mu$-term while $Y_{\alpha \beta \gamma}$ are the holomorphic Yukawa couplings. 

The gauge kinetic function for the visible sector localised at the singularity $\tau_\SM \rightarrow 0$ depends uniquely on the dilaton:
\be
f_a = \kappa_a S + \lambda_a T_\SM \rightarrow \kappa_a S \,,
\ee
where $\kappa_a$ is a singularity-dependent coefficient. 

\subsubsection{Moduli stabilisation}
\label{ssec:modulistabilisation}

Due to the no-scale structure of theories defined by a K\"ahler potential as in~\eqref{eq:kahlerpotential}, at tree level the supergravity F-term scalar potential is positive definite and depends only on $S$ and $U^\alpha$. The global supersymmetric minimum of such a scalar potential is located at:
\be
\label{eq:susystabilisation}
D_S W_0 = 0 \,, \qquad D_{U^\alpha} W_0 = 0 \,.
\ee
In LVS corrections due to $\alpha'$ effects in (\ref{eq:kahlerpotential}) and non-perturbative effects in (\ref{eq:superpotential}) are subleading with respect to (\ref{eq:susystabilisation}) in the large volume limit $\vo \gg 1$. Thus one can consistently first stabilise $S$ and $U^\alpha$ in a supersymmetric manner at order $\mc{O}\left(\vo^{-2}\right)$, and then fix the $T$-moduli by perturbing this solution with corrections at order $\mc{O}\left(\vo^{-3}\right)$. This perturbation generates also a small shift in the minimum \cite{Aparicio:2014wxa}: $D_S W \sim D_{U^\alpha} W \sim \mc{O}\left(\vo^{-1}\right)$ which is of fundamental importance to generate non-vanishing gaugino masses, as explained in Sec.~\ref{sec:soft-terms}. The low-energy F-term scalar potential of the LVS model described above takes the form:
\be
\label{eq:LVSscalarpotential}
V_F = \frac{g_s}{8 \pi} \left[\sum_{i = 1}^n \left(\frac{8}{3} (a_i A_i)^2 \sqrt{\tau_i} \,\frac{e^{-2 a_i \tau_i}}{\vo} -  4 a_i A_i W_0 \tau_i \,\frac{e^{-a_i \tau_i}}{\vo^2}\right) + \frac{3 \hat{\xi} |W_0|^2}{4 \vo^3}\right] + V_\dS \,,
\ee
where $V_\dS$ is a term which depends on the details of the dS sector. Notice that the first two terms descend from the non-perturbative effects in~\eqref{eq:superpotential}. The axions $\psi_i$, $i=1,...,n$, are stabilised in such a way that the second term is negative. The last term in square brackets is due to $\alpha'$ corrections in~\eqref{eq:kahlerpotential}. $V_\dS$ is essential to achieve a Minkowski/dS vacuum $V_0 \equiv \langle V \rangle \simeq 0$. Two dS sectors consistent with sequestered compactifications were considered in \cite{Aparicio:2014wxa}:
\ben
\item dS$_1$ case: \textit{de Sitter from hidden charged fields} \\ 
LVS string compactifications generically feature hidden sector D-brane stacks on the `big' divisor $D_b$ which support matter fields $\phi_\dS$ that are charged under a $U(1)$ symmetry \cite{CYembedding}. D-term stabilisation induces non-zero F-terms for these hidden sector fields which give rise to a positive definite contribution to the scalar potential, $V_\dS \sim W_0^2/\vo^{8/3}$, that can lead to Minkowski/dS vacua. From a higher-dimensional point of view, this corresponds to having a T-brane background \cite{Cicoli:2015ylx}.

\item dS$_2$ case: \textit{de Sitter from non-perturbative effects} \\ 
It is possible to have an additional shrinkable divisor $D_\dS$ on top of the O7-plane and supporting non-perturbative effects \cite{Cicoli:2012fh}. D-term stabilisation fixes the size of this divisor to $\tau_\dS = 0$,\footnote{Since $\tau_{\rm dS} \rightarrow 0$ the expression for the volume in~\eqref{eq:volume} is not affected by the new divisor $D_\dS$.} while a new term in the superpotential of the form $W_\dS = A_\dS \,e^{-a_\dS (S + \kappa_\dS T_\dS)}$ gives rise to a positive definite contribution to the F-term scalar potential with scaling $e^{-2 a_\dS \text{Re} (S)}/\vo$ which can lead to a Minkowski/dS vacuum.
\een

The position of the minimum for the scalar potential in (\ref{eq:LVSscalarpotential}) does not depend on the dS sector at leading order. It is given by the following relations:
\be
\label{eq:LVSminimum}
e^{-a_i \tau_i} \simeq  \frac{3 W_0 \sqrt{\tau_i}}{4 a_i A_i \vo}
 \, \qquad \text{and} \qquad \left(a_i\tau_i\right)^{3/2} \simeq \frac{\hat\xi}{2}\,\frac{1}{\sum_{i=1}^n a_i^{-3/2}} \quad\forall i=1,...,n\,.
\ee
These relations clearly imply that at the minimum $a_i \tau_i \sim \mc{O}\left(\ln \vo\right)$.

\subsubsection{F-terms}
\label{ssec:susybreaking}

SUSY-breaking is governed by F- and D-terms. However, in the setup of Sec.~\ref{sec:setup} D-terms are subdominant with respect to F-terms \cite{Aparicio:2014wxa} whose general expression in supergravity is: 
\be
F^i = e^{K/2} K^{i \overline{j}} D_{\overline{j}} \overline{W} \,.
\ee
Once supersymmetry is broken, the gravitino acquires a mass given by:
\be
m_{3/2} = e^{K/2} |W| \simeq \frac{g_s^{1/2}}{2 \sqrt{2 \pi}} \frac{W_0 M_p}{\vo}\,.
\ee
The major source of SUSY-breaking is the presence of background fluxes which generate a non-vanishing F-term for the field $\tau_b$ through a non-zero $W_0$. At leading order it does not depend on the dS sector and it is given by three terms which come from tree-level, $\alpha'$ and non-perturbative effects:
\be
F^{T_b} = F^{T_b}_{\rm tree} + F^{T_b}_{\alpha'} + F^{T_b}_{\rm np}\,,
\ee
where (the axions have already been fixed at their minimum):
\be
\frac{F^{T_b}_{\rm tree}}{m_{3/2}} = - 2 \tau_b\,,\qquad \frac{F^{T_b}_{\alpha'}}{m_{3/2}} = - \frac{3\tau_b}{2} \frac{\hat\xi}{\vo} \,,
\qquad \frac{F^{T_b}_{\rm np}}{m_{3/2}} = \sum_{i=1}^n\frac{4 A_i a_i \tau_i}{\sqrt{\tau_b}}\,\frac{\vo}{W_0}\,e^{-a_i\tau_i}\,.
\label{FTb}
\ee
The minimisation condition (\ref{eq:LVSminimum}) implies that at the minimum: 
\be
\langle F^{T_b}_{\rm np} \rangle = - \langle F^{T_b}_{\alpha'} \rangle + \mc{O}\left( \frac{\tau_b \,\hat\xi\,m_{3/2}}{\vo \ln\vo}\right).
\label{relFTb}
\ee
Thus at the minimum $F^{T_b}$ scales as: 
\be
\langle F^{T_b} \rangle = - 2 \tau_b m_{3/2}\left[1+ \mc{O}\left(\frac{\hat\xi}{\vo \ln\vo}\right)\right]\,.
\label{FTbmin}
\ee
Similarly to $F^{T_b}$, the F-terms associated to the `small' divisors $D_i$ receive also three contributions:
\be
F^{T_i} = F^{T_i}_{\rm tree} + F^{T_i}_{\alpha'} + F^{T_i}_{\rm np}\,,
\ee
where:
\be
\frac{F^{T_i}_{\rm tree}}{m_{3/2}} = - 2 \tau_i\,,\qquad \frac{F^{T_i}_{\alpha'}}{m_{3/2}} = - \frac{3\tau_i}{2} \frac{\hat\xi}{\vo} \,,
\qquad \frac{F^{T_i}_{\rm np}}{m_{3/2}} = \frac{8 A_i a_i \sqrt{\tau_i}}{3}\,\frac{\vo}{W_0}\,e^{-a_i\tau_i}\,.
\label{eq:smallfterms2}
\ee
Using (\ref{eq:LVSminimum}) it turns out that at the minimum: 
\be
\langle F^{T_i}_{\rm np} \rangle = - \langle F^{T_i}_{\rm tree} \rangle + \mc{O}\left( \frac{\tau_i m_{3/2}}{\ln\vo}\right).
\label{relFTi}
\ee
Hence at the minimum $F^{T_i}$ is determined by the first correction to $F^{T_i}_{\rm np}$ in the $\ln \vo$-expansion:
\be
\langle F^{T_i}\rangle \simeq - \frac{3 m_{3/2}}{2 a_i} \,.
\label{FTimin}
\ee
Moreover the scalar potential (\ref{eq:LVSscalarpotential}) can be rewritten as:
\be
V = \sum_{I,J \in \{b,i\}} K_{I \overline{J}} F^I \overline{F}^{\overline{J}} - 3 m_{3/2}^2 \,.
\ee
Due to the no-scale structure:
\be
\sum_{I,J \in \{b,i\}} K^0_{I \overline{J}} F^I_{\rm tree} \overline{F}^{\overline{J}}_{\rm tree} - 3\,m_{3/2}^2 = 0\,,
\ee
where $K^0_{I \overline{J}}$ is the tree-leve K\"ahler metric, the scalar potential (\ref{eq:LVSscalarpotential}) is generated by non-perturbative and $\alpha'$ corrections to the effective action. In detail, for each $i=1,...,n$, the three terms in square brackets in~\eqref{eq:LVSscalarpotential} are given by:
\ben
\item $\frac{g_s (a_i A_i)^2 \sqrt{\tau_i}}{3 \pi} \frac{e^{-2 a_i \tau_i}}{\vo} = K^0_{T_i \overline{T}_i} F^{T_i}_{\rm np} \overline{F}^{\overline{T}_i}_{\rm np}$\,,

\item $- \frac{g_s (a_i A_i) W_0 \tau_i}{2 \pi} \frac{e^{-a_i \tau_i}}{\vo^2}= \left(K^0_{T_b \overline{T}_i} F^{T_b}_{\rm tree} + K^0_{T_i \overline{T}_i} F^{T_i}_{\rm tree} \right) \overline{F}^{\overline{T}_i}_{\rm np}+ K^0_{T_b \overline{T}_b} F^{T_b}_{\rm tree} \overline{F}^{\overline{T}_b}_{\rm np} + \text{h.c.}$\,,

\item $\frac{3 g_s}{32 \pi} \frac{\hat\xi m_{3/2}^2}{\vo} = \left(K^0_{T_b \overline{T}_b} F^{T_b}_{\rm tree} \overline{F}^{\overline{T}_b}_{\alpha'} + \text{h.c.}\right) + K^{\alpha'}_{T_b \overline{T}_b} F^{T_b}_{\rm tree}\overline{F}^{\overline{T}_b}_{\rm tree}$\,,
\een
where $K^{\alpha'}_{T_b \overline{T}_b}$ is the leading order $\alpha'$ correction to the $T_b \overline{T}_b$ element of the K\"ahler metric.

Let us finally point out that the F-terms of the dilaton and the complex structure moduli are subleading relative to the others given that they are generated by the shift of their minimum induced by non-perturbative and $\alpha'$ corrections. This shift can be parameterised as in \cite{Aparicio:2014wxa}:\footnote{We report here only the leading order term in the $\ln\vo$-expansion. Subleading corrections are nevertheless important for phenomenological applications, as explained in Sec.~\ref{sssec:gauginomasses}.}
\be
F^S = s \,\omega(U,S) \frac{\tau_i^{3/2} m_{3/2}}{\vo} \, \qquad \text{and} \qquad \, F^{U^{\alpha}} = \beta^\alpha(U,S) F^S \,,
\ee
where $\omega(U,S)$ and $\beta^\alpha(U,S)$ are $\mc{O}(1)$ flux-dependent coefficients which can be tuned to produce interesting phenomenological results. As we will see in Sec.~\ref{sssec:gauginomasses}, even if $F^S$ is $\vo$-suppressed, it plays a key r\^ole in the phenomenology of sequestered models.

\subsection{Soft-terms}
\label{sec:soft-terms}

In sequestered string compactifications the visible sector lives on D3-branes at singularities \cite{SingD3s, CYembedding}. Due to this particular D-brane configuration, in sequestered models gauge degrees of freedom living on D3-branes decouple from bulk fields. Furthermore, since SUSY is broken in the bulk, this decoupling results in a suppression of soft-terms with respect to the gravitino mass $m_{3/2}$. As we shall explain in Sec.~\ref{sec:dynamics}, due to this separation of scales, sequestering can help to get low-energy SUSY without any cosmological problem associated to light moduli \cite{CMP} or gravitino overproduction \cite{gravProbl}. The aim of this section is to summarise the pattern that the soft terms acquire in sequestered models, depending on the form of the K\"ahler matter metric $\tilde{K}_\alpha$ in~\eqref{eq:kahlermattermetric1}.

\subsubsection{Gaugino masses}
\label{sssec:gauginomasses}

Given that the gauge kinetic function for D3-branes at singularities depends on the dilaton $S$, gaugino masses are controlled by $F^S$ and look like: 
\be
\label{eq:gauginomasses}
M = \frac{F^S}{2s} \simeq \lambda(U,S) \frac{\hat\xi \,m_{3/2}}{\vo}\,,
\ee
where $\lambda(U,S)$ is a flux-dependent function whose explicit expression depends on the dS sector and is given in \cite{Aparicio:2014wxa}.

\subsubsection{Scalar masses}
\label{sssec:scalarmasses}

The situation is much more involved for scalar masses whose general expression, assuming a diagonal K\"ahler metric $\tilde{K}_{\alpha \overline{\beta}} = \tilde{K}_\alpha \delta_{\alpha \overline{\beta}}$, is given by:
\be
m_\alpha^2 = m_{3/2}^2 + V_0 - F^I \overline{F}^{\bar{J}} \partial_I \partial_{\bar{J}} \ln \tilde{K}_\alpha + \tilde{K}_\alpha^{-1} \sum_a g_a^2 D_a \partial_\alpha \partial_{\overline{\alpha}} D_a \,.
\label{eq:scalarmassesgeneral}
\ee
Non-zero D-term contributions can arise from hidden scalar fields $\phi_k$ living on D7-branes wrapped around the large 4-cycle $\tau_b$ which supports an anomalous $U(1)$ with $g_b^2 = \tau_b^{-1}$. These D-term contributions look like:
\be
D_b = \sum_k Q_{k,b} \phi_k \frac{\partial K}{\partial \phi_k} + \sum_I q_{I,b} \partial_{T_I} K \,,
\ee
where $Q_{k,b}$ and $q_{I,b}$ are, respectively, the $U(1)$ charge of the $k$-th hidden scalar and the $I$-th K\"ahler modulus under the $U(1)$ on $D_b$. 

In order to compute scalar masses, it is crucial to know the exact moduli dependence of $\tilde{K}_\alpha$. The leading order results are (see \cite{Aparicio:2014wxa} for the details of the computation):
\ben
\item \textit{Ultra-local limit} \\
Due to the form of $\tilde{K}_\alpha$ in~\eqref{eq:uldefinition}, the $\mc{O}\left(\vo^{-3}\right)$ F-term contributions to scalar masses cancel off and the final result depends on the dS sector:
\bi
\item[a)] dS$_1$ case: non-zero scalar masses are generated by D-term contributions at $\mc{O}\left(\vo^{-3}\right)$ \cite{Aparicio:2014wxa}:
\be
m_0^2 \simeq \frac{9}{64} \frac{\hat\xi \,m_{3/2}^2}{\vo\ln\vo} \,,
\ee
while the leading F-term contribution to $m_0^2$ is at $\mc{O}\left(\vo^{-4}\right)$. In this case scalar masses are universal.

\item[b)] dS$_2$ case: D-term contributions are subleading with respect to $\mc{O}\left(\vo^{-4}\right)$ contributions from F-terms of $S$ and $U$-moduli which give:
\be
m_\alpha^2 \simeq Q_\alpha(U,S)\, M^2 \,,
\label{eq:scalarmassesdilaton}
\ee
where $Q_\alpha(U,S)$ is a flux-dependent function involving derivatives of $f_\alpha (U,S)$. In this specific case scalar masses might not be universal.
\ei

\item \textit{Local limit} \\
In the local limit the effect of D-terms is negligible. We report the results for two limiting cases:
\bi
\item[(a)] \textit{$c_n = - 1/3$}:
\be
m_0^2 \simeq \frac{15}{4} \left(c_\xi - \frac 13\right) \frac{\hat\xi\, m_{3/2}^2}{\vo} \,,
\label{Locala}
\ee

\item[(b)] \textit{$c_\xi=0$}:
\be
m_0^2 \simeq \frac{15}{4\,n} \left[c_n- \frac 13 \left(n-1\right)\right] \frac{\hat\xi\, m_{3/2}^2}{\vo} \,,
\label{eq:scalarmassesinflaton}
\ee
\ei
which implies that in the local limit scalar masses have to be universal. This result has been derived assuming, without loss of generality, $a_i=a$ $\forall i=1,...,n$, so that $\tau_i^{3/2} = \hat\xi /(2\,n)$.
\een

Vanishing leading order results for the scalar masses in the ultra-local limit can be recovered by setting $c_\xi=1/3$ in case (a) and $c_n= (n-1)/3$ in case (b). We stress that in case (b), given that $c_\xi=0$, an effective ultra-local limit can be obtained only at the minimum using the minimisation condition (\ref{eq:LVSminimum}). Notice that in all cases, except for the ultra-local dS$_1$ case, scalars can be either tachyonic or non-tachyonic, depending on $Q_{\alpha}(U,S)$, $c_n$ and $c_\xi$. Non-tachyonic scalars require $Q_\alpha(U,S)>0$ in the ultra-local dS$_2$ case and $c_\xi>1/3$ for case (a) and $c_n > (n-1)/3$ for case (b) of the local limit.

\subsubsection{$A$-terms}
\label{sssec:aterms}

The scalar potential (\ref{eq:LVSscalarpotential}) contains cubic terms in the canonically normalised scalar fields ${\hat C}^\alpha$ of the form:\footnote{Higher-order superpotential terms that lift $\phi$ result in $A$-terms that are higher than cubic, corresponding to $n > 3$ in (\ref{pot}). Here we perform an explicit computation of the cubic $A$-terms. Calculations are more involved for higher order $A$-terms but the results are qualitatively similar.}   
\be
V \supset \hat{Y}_{\alpha \beta \gamma} A_{\alpha \beta \gamma} \hat{C}^\alpha \hat{C}^\beta \hat{C}^\gamma\,,
\ee
where $A_{\alpha \beta \gamma}$ are functions of the moduli of the compactification. It turns out that D-term contributions to $A_{\alpha \beta \gamma}$ are subleading \cite{Aparicio:2014wxa}, while the general F-term expression for the $A$-terms is:
\be
\label{eq:generalaterms}
A_{\alpha \beta \gamma} = F^i \partial_i \left[K + \ln\left(\frac{Y_{\alpha \beta \gamma}(U,S)}{\tilde{K}_\alpha \tilde{K}_\beta \tilde{K}_\gamma}\right)\right] \,.
\ee
Writing the K\"ahler matter metric as $\tilde{K}_\alpha = f_\alpha \tilde{K}$, the general expression in~\eqref{eq:generalaterms} can be rewritten as:
\be
\label{eq:simplifiedaterms}
A_{\alpha \beta \gamma} = F^i \partial_i \left[K - 3 \ln \tilde{K} + \ln\left(\frac{Y_{\alpha \beta \gamma}(U,S)}{f_\alpha f_\beta f_\gamma}\right)\right].
\ee
We study separately the different limits of the K\"ahler matter metric:
\ben
\item \textit{Ultra-local limit}: \\
From~\eqref{eq:simplifiedaterms} it is straightforward to see that there is a cancellation between the first two terms in square brackets. As a consequence the $A$-terms are determined by the F-terms of the dilaton and the complex structure moduli:
\be
A_{\alpha \beta \gamma} = \sum_{i = S, U} F^i \partial_i \left[\ln\left(\frac{Y_{\alpha \beta \gamma}(U,S)}{f_\alpha f_\beta f_\gamma}\right)\right] \equiv \frac{\Pi(U,S)}{\vo^2}\,,
\ee
where $\Pi(U,S)$ is an $\mc{O}(1)$ flux-dependent function.

\item \textit{Local limit}: \\
In this case the $A$-terms receive contributions also from the F-term of $T_b$:
\bi
\item[(a)] $c_n=-1/3$:
\be
\label{eq:atermsdilaton}
A_{\alpha \beta \gamma} = \frac 92 \left(c_\xi - \frac 13\right) \frac{\hat\xi\,m_{3/2}}{\vo} + \frac{\Pi(U,S)}{\vo^2} \,,
\ee
where both terms have the same volume scaling.

\item[(b)] $c_\xi=0$:
\be
\label{eq:inflatonaterms}
A_{\alpha \beta \gamma} = \frac{9}{2\,n} \left[c_n- \frac 13 \left(n-1\right)\right]  \frac{\hat\xi\,m_{3/2}}{\vo} + \frac{\Pi(U,S)}{\vo^2}\,,
\ee
where again both terms have the same volume scaling.
\ei
\een

\subsubsection*{Summary of soft terms}
\label{sssec:summarysoftspectra}

Here we summarise the results of the last sections, noting that they can be divided in two classes depending on the form of the K\"ahler matter metric:
\ben
\item \textit{MSSM-like spectrum}: in the ultra-local dS$_2$ case all soft terms have the same volume scaling:
\be
M \sim m_\alpha \sim A_{\alpha \beta \gamma} \sim \frac{M_p}{\vo^2} \,.
\ee
This is a typical MSSM-like spectrum with (possibly non-universal) scalars and gauginos at the same energy scale.

\item \textit{Split SUSY spectrum}: in the local and ultra-local dS$_1$ cases the volume scaling of gauginos and scalars is different:
\be
M \sim A_{\alpha \beta \gamma} \sim \frac{M_p}{\vo^2} \, \qquad \text{while} \qquad m_0 \sim \frac{M_p}{\vo^{3/2}} \,.
\ee
Thus gauginos are lighter than scalars, featuring a splitting of the soft scales.
\een
Notice that in order to have TeV gaugino masses, the volume $\vo$ should be in the range $10^6$-$10^8$. In turn, in order for~\eqref{eq:LVSminimum} to be satisfied without the necessity to fine-tune the coefficients $A_i$, the following relation should be satisfied:
\be
a_i \tau_i \simeq \ln\left(\frac{\vo}{W_0}\right) \qquad \quad\forall i =1,...,n\,,
\label{eq:finetuningcondition}
\ee
where $\tau_i$ is given in terms of $g_s$ as in (\ref{eq:LVSminimum}). For typical values of $W_0$ in the range $1$-$100$ (\ref{eq:finetuningcondition}) implies that the ratio $a_i/g_s$ has to lie in the range $60$-$120$.\footnote{We will fix $\xi = 1$ and $n = 10$ to perform numerical calculations.}

\section{Inflation, reheating and Affleck-Dine baryogenesis}
\label{sec:dynamics}

As mentioned in Sec.~\ref{sec:ADreview}, in order to examine the viability of AD baryogenesis, one needs to have an explicit model for SUSY breaking and inflation. Here we use the model proposed in \cite{Conlon:2005jm} to realise inflation since in this inflationary scenario the range of $\vo$ required to get the observed amplitude of density perturbations leads also to low-energy gauginos. 

\subsection{Inflationary dynamics}
\label{ssec:KMI}

The idea behind the model proposed in \cite{Conlon:2005jm} is very simple: the `small' modulus $\tau_n$ plays the r\^ole of the inflaton which experiences an exponentially flat direction when moved away from its minimum. The CY volume $\vo$ is instead kept almost fixed during inflation by the additional `small' moduli $\tau_j$, $j=1,...,n-1$ which sit at their minima. Therefore the total scalar potential (\ref{eq:LVSscalarpotential}) during inflation takes the simplified form: 
\be
\label{eq:inflationarypotential1}
V = \frac{g_s}{8 \pi} \left[\sum_{j = 1}^{n-1} \left(\frac{8}{3} (a_j A_j)^2 \sqrt{\tau_j} \,\frac{e^{-2 a_j \tau_j}}{\vo} -  4 a_j A_j W_0 \tau_j \,\frac{e^{-a_j \tau_j}}{\vo^2}\right) + \frac{3 \hat{\xi} |W_0|^2}{4 \vo^3}\right] + V_\dS + \delta V (\tau_n ) \,.
\ee
where:
\be
\delta V(\tau_n ) = - \frac{g_s}{2 \pi}\,a_n A_n W_0 \tau_n \frac{e^{-a_n \tau_n}}{\vo^2}\,.
\ee
If all $\tau_j$, $j=1,...,n-1$ and $\vo$ are fixed at their minima during inflation, the potential in~\eqref{eq:inflationarypotential1} can be written as the sum of a constant $V_0$ and $\delta V (\tau_n )$:
\be
\label{eq:inflationarypotential2}
V_{\rm inf} = V_0 - \frac{g_s}{2 \pi}\,a_n A_n W_0 \tau_n \frac{e^{-a_n \tau_n}}{\vo^2} \,,
\ee 
where a careful computation gives:
\be
V_0 = \frac{3}{4\,n} \frac{\hat\xi\, m_{3/2}^2}{\vo}\,.
\label{eq:v0}
\ee
We stress that the minima of $\tau_j$, $j=1,...,n-1$ and $\vo$ during inflation are slightly shifted from their values after the end of inflation:
\be
\left.\ln\vo\right|_{\rm inf} \underset{n\gg 1}{\simeq} \ln \vo \left(1+ \frac{2}{n}\right) \qquad\text{and}\qquad 
\left.\tau_j^{3/2}\right|_{\rm inf} \underset{n\gg 1}{\simeq} \tau_j^{3/2}\left( 1+\frac 3n\right).
\label{shift}
\ee
Given that the canonically normalised inflaton $\sigma$ is given by:
\be
\sigma = \sqrt{\frac{4}{3 \vo}} \,\tau_n^{3/4}\,,
\label{eq:canonicallynormalizedinflaton}
\ee 
the scalar potential for $\sigma$ takes the form:
\be
V_{\rm inf} = V_0 - c_1 \sigma^{4/3} e^{-c_2 \sigma^{4/3}}   \qquad\text{with}\quad c_1 = \frac{g_s}{2 \pi} \frac{A_n W_0}{\vo^2} c_2  \quad\text{and}\quad
c_2= a_n \left(\frac{3 \vo}{4}\right)^{2/3}\,.
\label{eq:inflationarypotential3}
\ee 
It is much easier to express the slow-roll parameters in terms of the non-canonically normalised fields as:
\bea
&\epsilon& \simeq \frac{512 \, n^2}{27} \frac{a_n^3 A_n \tau_n^{5/2}}{\hat{\xi}^2 W_0^2}\, \vo^3 \, e^{-2 a_n \tau_n}\,, \\
&\eta& \simeq -\frac{64 \, n}{9} \frac{a_n^3 A_n \tau_n^{3/2}}{\hat\xi W_0}\, \vo^2\,  e^{- a_n \tau_n}\,.
\eea
In order for inflation to take place, both $\epsilon$ and $\eta$ have to be much smaller than $1$. Just looking at the volume scaling, and since in the late-time minimum $a_n \tau_n \sim \ln \vo$, it is easy to infer that inflation can take place in the region:
\be
a_n \tau_n \gtrsim 2 \ln \vo\,,
\ee
and it ends at $\tau_n^{\rm end}$ when $\epsilon$ becomes of order $1$. Moreover we have been able to numerically compute the position of the inflaton $\tau_n^*$ corresponding to horizon exit:
\be
N_e(\tau_n^*) \simeq \frac{9}{64 \, n} \frac{\hat\xi W_0}{a_n^2 A_n \vo^2} \int_{\tau_n^{\rm end}}^{\tau_n^*} \frac{e^{a_n \tau_n}}{\tau_n^{3/2}} \,d \tau_n \simeq 60\,,
\ee
where $N_e$ denotes the number of e-foldings. Once $\tau_{\rm inf}^*$ is known, both the tensor-to-scalar ratio $r$ and the spectral index $n_s$ can be evaluated as:
\be
r = 16 \epsilon(\tau_n^*) \qquad \text{and} \qquad n_s = 1 - 6 \epsilon(\tau_n^*) + 2 \eta(\tau_n^*) \,.
\ee
Typical values are $n_s \simeq 0.967$ and $r \lesssim 10^{-10}$, reflecting the small-field nature of this inflationary model.
Finally, it is necessary to numerically evaluate the amplitude of density perturbations $\mc{A}_{\mc{R}}$ as a function of $\tau_n$, and to impose that it matches the measured value at $\tau_n^*$. This requirement translates into:
\be
\label{eq:COBE}
\mc{A}^2_{\mc{R}}(\tau_n^*) \simeq \frac{g_s}{8 \pi} \frac{3^4 \hat\xi^3 W_0^4}{4^6 n^3 \, a_n^4 A_n^2 (\tau_n^*)^{5/2}} \frac{e^{2 a_n \tau_n^*}}{\vo^6} \simeq 2.7 \times 10^{-7}\,.
\ee
The numerical analysis proceeds as follows: we set $a_j=a_n$ and $A_j=A_n$ $\forall j=1,...,n-1$ and choose the values of $a_n$ and $g_s$ as explained around (\ref{eq:finetuningcondition}) in order to avoid a severe fine-tuning of $A_n$ due to the relation (\ref{eq:LVSminimum}). Then for different values of $W_0$ in the natural range $1$-$100$ we compute the value of the volume $\left.\vo\right|_{\scriptscriptstyle \rm COBE}$ which reproduces the measured amplitude of density perturbations using (\ref{eq:COBE}). In Tab.~\eqref{tab:tab1} we report the results for $a_n = 2 \pi$, $g_s = 0.06$ and $n = 10$.

\begin{table}[h!]
\centering
\begin{tabular}{cccc}
\hline
 & $W_0$ & $\left.\log_{10}\left(\vo\right|_{\scriptscriptstyle \rm COBE}\right)$ & $A_n$\\
\hline
$(A)$ & $1$ & $5.14$ & $1.94$ \\
\hline
$(B)$ & $10$ & $6.15$ & $1.91$ \\
\hline
$(C)$ & $100$ & $7.15$ & $1.87$ \\
\hline
\end{tabular}
\caption{Values of the volume $\left.\vo\right|_{\scriptscriptstyle \rm COBE}$ which match the observed amplitude of density perturbations for different values of $W_0$ and $a_n = 2 \pi$, $g_s = 0.06$ and $n = 10$.}
\label{tab:tab1}
\end{table}

\subsection{Dynamics of the Affleck-Dine field}
\label{ssec:ADBdynamics}

The dynamics of the AD field $\phi$ is governed by the scalar potential (\ref{pot}). In order to determine whether $\phi$ can acquire a large VEV during inflation, we need to compute its soft mass and the corresponding $A$-term. Since $\phi$ is made of MSSM scalars, we can readily use the expressions derived in Sec.~\ref{sssec:scalarmasses} and Sec.~\ref{sssec:aterms}. In what follows, we shall analyse separately the MSSM-like and the split SUSY case. Below, we denote with $\tilde{x}$ the value of any quantity $x$ during inflation. 

\subsubsection{MSSM-like case}
\label{sssec:ADdynamicsMSSM}

As explained in Sec.~\ref{sssec:scalarmasses}, an MSSM-like spectrum arises in the ultra-local dS$_2$ case where D-terms are negligible. Using (\ref{eq:scalarmassesgeneral}) and $V_0 = |F|^2 - 3 m_{3/2}^2$, due to the ultra-local condition (\ref{eq:uldefinition}), the soft mass of $\phi$ becomes:
\be
\tilde{m}_\phi^2 = m_{3/2}^2 + V_0 - \frac{1}{3} K_{I\bar{J}} F^I \overline{F}^{\bar{J}} -  F^I \overline{F}^{\bar{J}} \partial_I \partial_{\bar{J}} \ln f_\alpha(U,S) = \frac 23 \,V_0 + Q_\alpha(U,S) \,M^2\,.
\ee
Given that $V_0 \sim \mc{O}\left(\vo^{-3}\right)$ while $M^2\sim\mc{O}\left(\vo^{-4}\right)$, the leading order contribution to the soft mass of $\phi$ during inflation comes from the vacuum energy $V_0$: 
\be
\tilde{m}^2_\phi \simeq \frac 23 V_0 =  \frac{1}{2\,n} \frac{\hat{\xi} m_{3/2}^2}{\vo} \simeq 2 H^2_{\rm inf}> 0\,.
\label{mphiUL}
\ee
This implies that in this case the AD field cannot acquire a tachyonic mass during inflation. Thus $\phi$ settles at the origin during inflation and remains there throughout the entire post-inflationary history. As a result, it cannot give rise to a successful AD baryogenesis. This is in agreement with similar results found in \cite{Casas:1997uk, Dutta:2010sg, Marsh:2011ud}.

\subsubsection{Split SUSY case}
\label{sssec:ADdynamicssplitsusy}

During inflation the inflaton $\tau_n$ is displaced from its late-time minimum. Therefore the inflaton-dependent contribution to $F^{T_b}_{\rm np}$ in (\ref{FTb}) and to $F^{T_n}_{\rm np}$ in (\ref{eq:smallfterms2}) can be neglected during inflation, leading to:
\be 
\frac{\tilde{F}^{T_b}_{\rm np}}{m_{3/2}} = \sum_{j=1}^{n-1}\frac{4 A_j a_j \tau_j}{\sqrt{\tau_b}}\,\frac{\vo}{W_0}\,e^{-a_j\tau_j}\,,
\ee
and:
\be
\frac{\tilde{F}^{T_n}_{\rm np}}{m_{3/2}} = \frac{8 A_n a_n \sqrt{\tau_n}}{3}\,\frac{\vo}{W_0}\,e^{-a_n\tau_n}  \rightarrow 0\,.
\ee
Thus the relations (\ref{relFTb}) and (\ref{relFTi}) for $i=n$, which were true at the minimum, do not hold anymore during inflation when the F-terms of $T_b$ and $T_n$ become:
\be
\tilde{F}^{T_b} = -2\tau_b\,m_{3/2}\left(1+\frac{3}{4\,n} \frac{\hat\xi}{\vo}\right)\,,
\ee
and:
\be
\tilde{F}^{T_n} = \tilde{F}^{T_n}_{\rm tree} = - 2 \tau_n m_{3/2} \,.
\ee
Given that both $F^{T_b}$ and $F^{T_n}$ have a different form during inflation compared with the one at the minimum (see (\ref{FTbmin}) and (\ref{FTimin}) for $i=n$ respectively), in the local case, which leads to split SUSY, the AD field can possibly develop a tachyonic mass during inflation. 
In order to investigate this possibility, we consider the two limiting cases studied before:
\bi
\item[(a)] $c_n=-1/3$ \\
The scalar masses are determined by the following contributions:
\ben
\item $\tilde{F}^{T_b} \tilde{F}^{\overline{T}_b} \partial_{T_b} \partial_{\overline{T}_b} \ln\tilde{K}_\alpha = m_{3/2}^2 + \frac{5}{2} \frac{m_{3/2}^2}{\vo} \,\tau_n^{3/2} - \frac{15}{4} \frac{\hat\xi m_{3/2}^2}{\vo} \left(c_\xi - \frac 13 -\frac{1}{15 n}\right)$\,,
\item $\tilde{F}^{T_n} \tilde{F}^{\overline{T}_b} \partial_{T_n} \partial_{\overline{T}_b} \ln\tilde{K}_\alpha 
       + \tilde{F}^{T_b} \tilde{F}^{\overline{T}_n} \partial_{T_b} \partial_{\overline{T}_n} \ln\tilde{K}_\alpha = -3 \frac{m_{3/2}^2}{\vo}\,\tau_n^{3/2}$\,,
\item $\tilde{F}^{T_n} \tilde{F}^{\overline{T}_n} \partial_{T_n} \partial_{\overline{T}_n} \ln\tilde{K}_\alpha = \frac{1}{2} \frac{m_{3/2}^2}{\vo}\,\tau_n^{3/2}$\,.
\een
Using the general expression for scalar masses (\ref{eq:scalarmassesgeneral}), we find that the contributions proportional to $\tau_n^{3/2}$ cancel off giving a soft mass of $\phi$ during inflation of the form:
\be
\tilde{m}_\phi^2 = \frac{15}{4}\frac{\hat\xi m_{3/2}^2}{\vo} \left(c_\xi - \frac 13 + \frac{2}{15\,n}\right)\,.
\label{mphi}
\ee 
Notice that this expression correctly reproduces the soft scalar mass at the minimum (\ref{Locala}) in the limit $n \rightarrow \infty$, and the result (\ref{mphiUL}) in the ultra-local limit $c_\xi=1/3$. Moreover (\ref{mphi}) becomes negative if:
\be
c_\xi < \frac 13 - \frac{2}{15\,n}\,.
\ee
Given that this condition is in clear contrast with the requirement of non-tachyonic scalar masses after the end of inflation, i.e. $c_\xi > \frac 13$, we conclude that in this case it is not possible to obtain a tachyonic mass for the AD field during inflation.\footnote{The $A$-terms take the same form both during and after inflation.}

\item[(b)] $c_\xi=0$ \\
The scalar masses are determined by the following contributions:
\ben
\item $\tilde{F}^{T_b} \tilde{F}^{\overline{T}_b} \partial_{T_b} \partial_{\overline{T}_b} \tilde{K}_\alpha = m_{3/2}^2 - \frac{15}{2} c_n \frac{m_{3/2}^2}{\vo}\,\tau_n^{3/2} + \frac 54 \frac{\hat\xi m_{3/2}^2}{\vo}\left(1+\frac{1}{5\,n}\right)$\,,
\item $\tilde{F}^{T_n} \tilde{F}^{\overline{T}_b} \partial_{T_n} \partial_{\overline{T}_b} \tilde{K}_\alpha 
+\tilde{F}^{T_b} \tilde{F}^{\overline{T}_n} \partial_{T_b} \partial_{\overline{T}_n} \tilde{K}_\alpha= 9 c_n \frac{m_{3/2}^2}{\vo}\,\tau_n^{3/2}$\,,
\item $\tilde{F}^{T_n} \tilde{F}^{\overline{T}_n} \partial_{T_n} \partial_{\overline{T}_n} \tilde{K}_\alpha = -\frac 32 c_n \frac{m_{3/2}^2}{\vo}\,\tau_n^{3/2}$\,.
\een
Again the contributions proportional to $\tau_n^{3/2}$ cancel off in the general expression for scalar masses (\ref{eq:scalarmassesgeneral}), leading to a soft mass of $\phi$ during inflation of the form:
\be
\tilde{m}_\phi^2 = - \frac 54 \frac{\hat\xi m_{3/2}^2}{\vo} \left(1-\frac{2}{5 n}\right)\,,
\ee
which is always negative for $n>1$. Therefore the case with $c_\xi=0$ guarantees that the mass of the AD field becomes tachyonic during inflation for every value of $c_n$. On the other hand, as we have seen in Sec. \ref{sssec:scalarmasses}, if $c_n > (n-1)/3$, $m^2_\phi > 0$ after the end of inflation. Thus this case represents a good example where AD baryogenesis can be explicitly realised. We finally mention that also the $A$-terms get modified during inflation since they look like:
\be
A = - \frac 32 \frac{\hat\xi m_{3/2}}{\vo}\,.
\ee
\ei

\subsection{Reheating from lightest modulus decay}

In order to understand how reheating takes place we need first to look at the moduli mass spectrum. The canonically normalised inflaton $\sigma$ is exponentially light during inflation but after inflation, when it oscillates around its minimum, it becomes very heavy:
\be
m_\sigma \sim \frac{W_0 M_p}{\vo}\,\ln\vo \sim  m_{3/2}\,\ln\vo\,.
\label{msigma}
\ee
Due to the local nature of this blow-up mode, $\sigma$ is coupled to the field theory living on this 4-cycle as $1/M_s \sim \sqrt{\vo}/M_p$, as opposed to a standard Planckian-strength coupling. It therefore decays relatively quickly leading to an initial reheating temperature of order \cite{Cicoli:2010ha}: 
\be
T_{\rm rh,in} \sim \sqrt{\Gamma_{\sigma} M_p} \sim \frac{m_\sigma}{M_s} \sqrt{m_\sigma M_p} \sim \frac{M_p}{\vo}\,.
\label{Trhin}
\ee
If $\tau_n$ supports a hidden sector, when $\sigma$ decays, the inflaton will dump most of its energy to hidden sector degrees of freedom. This is not necessarily a problem since these hidden degrees of freedom get diluted by the decay of the lightest modulus. On the other hand, one has to check that the reheating temperature due to the decay of the inflaton does not give rise to thermal effects which destabilise the zero-temperature minimum. In LVS models this requires \cite{Anguelova:2009ht}:
\be
T_{\rm rh,in} < T_{\rm max} \sim \frac{M_p}{\vo^{3/4}}\,,
\ee
which is safely satisfied for $\vo \gg 1$ by the reheating temperature given in (\ref{Trhin}).

As shown in (\ref{shift}), during inflation all the blow-up modes $\tau_j$, $j=1,...,n-1$ get shifted from their late-time minimum. However they have a mass of order (\ref{msigma}) which makes them heavier than the inflationary Hubble scale given in (\ref{eq:v0}) since:
\be
\left(\frac{ m_{\sigma_j}}{H_{\rm inf}}\right)^2 \sim \frac{\vo}{\sqrt{\ln\vo}} \gg 1\,.
\ee
Therefore, after the end of inflation, these fields quickly relax to their late-time minimum and do not play any relevant r\^ole for the post-inflationary evolution. 

The situation is completely different for $\tau_b$ which is the lightest modulus. In fact, the canonically normalised `big' modulus $\chi$ acquires a mass whose exact form depends on the way to achieve a dS vacuum \cite{Aparicio:2014wxa}: 
\be
{\rm dS}_1: \quad m_\chi^2 \simeq \frac{9}{16 a_n \tau_n} \frac{\hat\xi \,m_{3/2}^2}{\vo}\,, \qquad \qquad 
{\rm dS}_2: \quad m_\chi^2 \simeq \frac{27}{8 a_n \tau_n} \frac{\hat\xi\, m_{3/2}^2}{\vo} \,,
\label{eq:volumemass}
\ee
implying that $\chi$ has a mass of order $H$ during inflation. Moreover, as shown in (\ref{shift}), $\chi$ gets a displacement in Planck units during inflation of order:
\be
\chi = \sqrt{\frac 23}\ln\vo\quad\Rightarrow\quad \Delta\chi = \chi_{\rm inf} -\chi \underset{n\gg 1}{\simeq} \frac{2\chi}{n} =
\frac{2}{n} \sqrt{\frac 23}\ln\vo \sim \mc{O}(1)\,.
\ee
Thus right after inflation $\chi$ starts oscillating around its minimum with an initial amplitude of order $M_p$. Being only gravitationally coupled to other fields, $\chi$ decays very late when it dominates the energy density of the universe, leading to dilution of any previously produced relic abundance and a final reheating temperature \cite{DMDRcorr}:
\be \label{eq:treheatingmod}
T_{\rm rh} \simeq \frac{0.3}{g_*^{1/4} \sqrt{\Delta N_{\rm eff}}}\,m_\chi\sqrt{\frac{m_\chi}{M_p}}\,,
\ee
where $g_*$ is the number of relativistic degrees of freedom at reheating while $\Delta N_{\rm eff}$ is the amount of extra axionic dark radiation produced from the decay of the lightest modulus \cite{Cicoli:2012aq,Cicoli:2015bpq}. For $g_*\sim 100$ and $\Delta N_{\rm eff}\simeq 0.5$,\footnote{The actual prediction for axionic dark radiation is model-dependent: in MSSM-like scenarios the main visible sector decay channel for $\chi$ is into Higgses, while in split SUSY models $\phi$ can also decay into squarks and sleptons. For details, see \cite{Cicoli:2012aq,Cicoli:2015bpq}.} and using the numbers in Tab. \ref{tab:tab1}, we find $m_\chi \sim 10^{8}-10^{9}$ GeV and $T_{\rm rh}\sim 10^{3}-10^{4}$ GeV. This value of $T_{\rm rh}$ is larger than BBN temperatures of order $1$ MeV, and so this model does not suffer from any cosmological moduli problem \cite{CMP}. Moreover, since $m_\chi/m_{3/2} \sim \vo^{-1/2} \ll 1$, the decay of the lightest modulus into gravitinos is kinematically forbidden, implying the absence of moduli-induced gravitino problem \cite{gravProbl}.

\subsection{Generation of baryon asymmetry}
\label{ssec:baryonasymmetry}

We now compute the baryon asymmetry of the universe generated via the AD mechanism in our model. As mentioned earlier, the decay of the lightest modulus $\chi$ reheats the universe at late times and dilutes the entropy produced from the decay of the inflaton $\sigma$. We are therefore interested in deriving the ratio $n_\B/s$, where $n_\B$ is the baryon number density stored in the AD field and $s$ is the entropy density produced by the decay of $\chi$. The decay of $\phi$ takes place when its VEV is redshifted to a sufficiently low value such that $y |\phi| < m_\phi$ ($y$ being a gauge or Yukawa coupling), at which point the decay of $\phi$ to the fields that are coupled to it and acquire an induced mass $y |\phi|$, is kinematically allowed. We have checked that even for $\phi_0 \sim M_p$, the decay of the AD field occurs well before the decay of $\chi$, and hence any entropy that it may produce will be diluted by the decay of $\chi$. 

Let us focus on the split SUSY case where the AD field can acquire a tachyonic mass during inflation. In this case, $m_\phi \sim m_\chi \sqrt{\ln\vo} > m_\chi\sim  H_{\rm inf}$, implying that both $\phi$ and $\chi$ start oscillating straight after the end of inflation when the AD field makes a very rapid transition from the tachyonic to the non-tachyonic regime.\footnote{In cases where $m_\phi \ll H_{\rm inf}$, thermal effects from inflaton decay may lead to early oscillations of the AD field~\cite{thermal}. However, this is not an important effect in our model.} Hence, at the onset of $\chi$ oscillations, i.e. at $H \sim H_{\rm inf}$, we have:          
\be \label{ratio2}
{n_\phi \over n_\chi} \simeq {m_\phi \over m_\chi} \left({\phi_0 \over M_p}\right)^2\,, 
\ee
where we have used $n_\chi \simeq m_\chi M^2_p$ and $n_\phi \simeq m_\phi \phi^2_0$. 
The BAU generated via the AD mechanism is then given by:
\be 
{n_\B \over s} =  {n_\B \over n_\phi} {n_\phi \over n_\chi} {n_\chi \over s} ,
\ee
where:
\be
{n_\chi \over s} = {3 T_{\rm rh} \over 4 m_\chi}\,,
\ee
is the yield from $\chi$ decay and $T_{\rm rh}$ is the reheating temperature of the universe after $\chi$ decay. After using the expression (\ref{asym}), and for $n \theta_i \sim \mc{O}(1)$, we arrive at the final result for baryon asymmetry (see also \cite{Kane:2011ih} for a similar result):
\be \label{bau1}
{n_\B \over s} \sim {|A| \over m_\chi} {T_{\rm rh} \over m_\chi} \left({\phi_0 \over M_p}\right)^2\,.
\ee

\section{Results and discussion}
\label{SecResults}

\subsection{Numerical results}
\label{sssec:numericalresults}

In this Section, we numerically analyse the parameters of the model so that the following requirements are satisfied: 
\begin{itemize}
\item[a)] A successful model with around $60$ e-foldings of inflation that creates density perturbations of the correct size with
all underlying parameters in their natural range;

\item[b)] Low-energy gauginos which are well motivated by gauge coupling unification;

\item[c)] A correct generation of the observed BAU, $n_\B/s \simeq 10^{-10}$, via the AD mechanism.
\end{itemize}

Requirement $a)$ has already been studied in Sec. \ref{ssec:KMI}. For concreteness, we consider the points in the parameter space listed in Tab. \ref{tab:tab1}. Scalar masses are completely determined by requirement $a)$ since, as we have shown in (\ref{sssec:scalarmasses}), they depend only on $W_0$ and $\vo$ once the string coupling constant has been fixed. On the contrary, requirement $b)$ can be easily fulfilled using the additional freedom of gaugino masses: the possibility to vary $\lambda(U,S)$ in (\ref{eq:gauginomasses}) by tuning background fluxes. In particular we require that $M = 5 \times 10^3 \, \rm GeV$ when $W_0 = 100$ and the value of the volume is the corresponding $\left.\vo\right|_{\scriptscriptstyle \rm COBE}$ in case $(C)$ of Tab. \ref{tab:tab1}. In order for requirement $c)$ to be fulfilled, it is necessary to find the exact value of the AD field displacement $\phi_0/M_p$ such that the baryon asymmetry estimated in (\ref{bau1}) matches the measured value. We focus on the split SUSY case where late-time scalars are non-tachyonic while the AD field during inflation becomes tachyonic. Moreover we assume that the A-terms are determined solely by the first term in (\ref{eq:atermsdilaton}). We stress that these choices do not affect the qualitative behaviour of our final results. 

\begin{figure}[h!]
\begin{center}
\includegraphics[width=0.55\textwidth, angle=0]{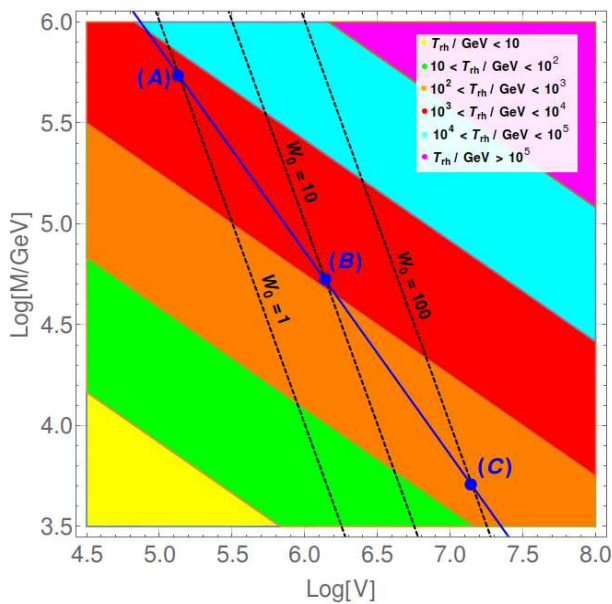}
\caption{$T_{\rm rh}$ as a function of $\vo$ and gaugino masses $M$ for $a_i=2\pi$ and $g_s = 0.06$. The blue dots correspond to the points of the parameter space in Tab.~\ref{tab:tab1}. The amplitude of density perturbations in these points matches the measured one provided that the displacement of the AD field at the start of oscillations is that given in Tab.~\ref{tab:tab2}.} \label{fig:plot1}
\end{center}
\end{figure}

In Fig.~\ref{fig:plot1} we illustrate the correlation between the produced baryon asymmetry and the gaugino masses in the split SUSY case. For the sake of concreteness we choose to work in the dS$_1$ scenario but the results are not dramatically affected by changing the dS sector. Differently coloured bands in Fig.~\ref{fig:plot1} correspond to different ranges for the reheating temperature. The dotted black lines correspond to constant values of $W_0$ in the natural range: $1, 10, 100$ from left to right. The continuous blue line corresponds to the locus where the amplitude of the density perturbations matches the measured value. It intersects the dotted black lines in the blue dots which respectively correspond to the cases $(A), (B), (C)$ in Tab. \ref{tab:tab1}. In Tab.~\ref{tab:tab2} we report the values of scalar masses, which are around $10^{9-10}$ GeV, the reheating temperature, which is larger than $100$ GeV, and the displacement of the AD field. Notice that it is possible to satisfy the requirements $a)$, $b)$ and $c)$ for natural $\mc{O}(1)$ values of the parameter $A_n$ and for $\phi_0 \sim 0.1\,M_p$. Moreover $T_{\rm rh}$ shown in Tab.~\ref{tab:tab2} and given in~(\ref{eq:treheatingmod}) is completely determined by the requirement of getting the right amplitude of density perturbations.

\begin{table}[h!]
\begin{center}
\begin{tabular}{ccccc}
\hline
 & $M \, [\rm{GeV}]$ & $m_0 \, [\rm{GeV}]$ & $T_{\rm rh} \, [\rm{GeV}]$ & $\phi_0/M_p$\\
\hline
$(A)$ & $5.4 \times 10^5$ & $3 \times 10^{10}$ & $6.7 \times 10^3$ & $0.03$ \\
\hline
$(B)$ & $5.2 \times 10^4$ & $9.2 \times 10^9$ & $1.1 \times 10^3$ & $0.08$ \\
\hline
$(C)$ & $5 \times 10^3$ & $2.8 \times 10^9$ & $195$ & $0.19$ \\
\hline
\end{tabular}
\end{center}
\caption{Gaugino masses $M$, scalar masses $m_0$, reheating temperature $T_{\rm rh}$ and displacement $\phi_0/M_p$ needed to match the measured amplitude of density perturbations for the cases listed in Tab.~\ref{tab:tab1}.}
\label{tab:tab2}
\end{table}

\subsection{Origin of dark matter abundance}

In order to determine whether the DM relic abundance has a thermal or non-thermal origin, we have to compare $T_{\rm rh}$ with the DM freeze-out temperature $T_{\rm f} \sim m_\DM/20$. If $T_{\rm rh} \, > \, T_{\rm f}$, the DM content is set by thermal freeze-out while, for $T_{\rm rh} \, < \, T_{\rm f}$, the DM abundance is produced non-thermally from the decay of the lightest modulus. Since the lowest reheating temperature that is compatible with successful inflation and baryogenesis is $T_{\rm rh} \, \simeq \, 195$ GeV, the non-thermal mechanism requires $m_\DM \, > \, 3.9$ TeV. Due to the gravity-mediated pattern of gaugino masses, the lightest gaugino is the Bino, and hence the DM candidate in our model is either a Bino- or a Higgsino-like neutralino. Binos typically have a small annihilation rate, $\langle \sigma_{\rm ann} v \rangle \, < \, 3 \times 10^{-26}$ cm$^3$ s$^{-1}$, especially for a split SUSY spectrum because of the extremely heavy sparticles. For Higgsinos with a mass above $\simeq 1.2$ TeV, we also have $\langle \sigma_{\rm ann} v \rangle \, < \, 3 \times 10^{-26}$ cm$^3$ s$^{-1}$.

Because of the smallness of the annihilation rate, non-thermal DM production must proceed through the `Branching' scenario where the correct relic abundance is produced directly from the decay of the lightest modulus~\cite{NonThDMinSeqLVS}. In this scenario, the DM abundance is given by:
\be \label{prodabun}
\left({n_\DM \over s}\right)_{\rm non-th} = {3 T_{\rm rh} \over 4 m_\chi} ~ {\rm Br}_\DM \,,
\ee
where ${\rm Br}_\DM$ is the branching ratio for producing $R$-parity odd particles (which eventually decay to the DM particle) from modulus decay. Even allowing for ${\rm Br}_\DM \sim 10^{-3}$, which is the smallest value allowed in this scenario~\cite{ADS}, and after using the values in Tab. \ref{tab:tab2}, we find that the `Branching' scenario would lead to DM overproduction by few orders of magnitude above the observed value: 
\be \label{obsabun}
\left({n_\DM \over s}\right)_{\rm obs} \simeq 5 \times 10^{-10} ~ \left({1 ~ {\rm GeV} \over m_\DM}\right)\,.
\ee
This implies that non-thermal DM is not compatible with inflation and baryogenesis in this model. We are therefore forced to consider thermal Higgsino DM with $m_\DM \simeq 1.2$ TeV where thermal freeze-out can produce the right DM abundance. For $m_\DM < 1.2$ TeV, the Higgsino is thermally underproduced, and so we need to consider mixed DM, as in the axion-Higgsino scenario~\cite{Howie}.                

Regarding the production of dark radiation, it has recently been shown \cite{Cicoli:2015bpq} that split SUSY models arising in sequestered string compactifications do not feature any overproduction. This is due to the large suppression of the excess of the effective neutrino number $\Delta N_{\rm eff}$ coming from the decay of $\chi$ into MSSM scalars, which is allowed in a vast region of the parameter space.

Finally we would like to comment on obtaining a large VEV for the AD field and possible implications for the DM content of the universe. It is seen in Tab. \ref{tab:tab2} that the generation of the observed BAU needs $m_\phi \sim 10^9$-$10^{10}$ GeV and $\phi_0 \sim 0.1$ $M_p$. It is possible to get $\phi_0$ in this ballpark, see~(\ref{initial}), if the AD field is lifted by a non-renormalisable term of level $n = 9$ where $\lambda_9 \sim 1$.\footnote{In fact, all of the MSSM flat directions are lifted at this level if the superpotential includes all higher-order terms that are compatible with gauge symmetry~\cite{gkm}.} In this case, depending on the Higgsino mass, one can have either Higgsino or mixed DM scenario as mentioned above. 

One may also obtain the required value of $\phi_0$ if the AD field is lifted by a renormalisable term with $n = 3$. However, in this case, a very small coupling $\lambda_3 \sim 10^{-8}-10^{-7}$ is needed. This is much smaller than all of the SM Yukawa couplings, but it may arise from renormalisable superpotential terms that violate $R$-parity (namely $LLE$, $UDD$, $QLD$ terms). Such terms destabilise the Higgsino, and a question is whether this can lead to a cosmologically consistent scenario. To answer this, let us consider the situation in the presence of the $LLE$ term. In this case, the Higgsino can decay to three leptons via an off-shell slepton. The decay rate is $\Gamma_{\tilde H} \sim (\lambda_3 y_l)^2 m^5_{\tilde H}/(8 \pi \cdot 32 \pi^2) m^4_{\tilde l}$, where $y_l$ is a leptonic Yukawa coupling, $m_{\tilde H}$ and $m_{\tilde l}$ denote the Higgsino and slepton masses respectively, and the factor of $32 \pi^2$ arises due to the three-body final state. For $\lambda_3 \sim 10^{-7}$, $y_l \sim 10^{-2}$, $m_{\tilde l} \sim 10^{10}$ GeV, and $m_{\tilde H} \lsim 200$ GeV, we may find a decay lifetime $\tau_{\tilde H} \gsim 10^{27}$ sec. This is compatible with the tightest cosmological bounds on decaying DM from the cosmic microwave background~\cite{slatyer}. The DM content of the universe can be explained within a mixed scenario where the Higgsino is the sub-dominant component.                

\section{Conclusions}
\label{Concl}

In this paper, we have presented a successful embedding of AD baryogenesis in type IIB sequestered string models. AD baryogenesis is a suitable mechanism for generating the observed BAU at temperatures below the EW scale. Such relatively low temperatures typically arise from the late-time decay of long-lived moduli in string compactifications. However, an explicit embedding of AD baryogenesis in such models is non-trivial as supergravity corrections can ruin the flatness of the AD field that is essential for the success of the AD mechanism.

The model presented here is in the context of LVS models and describes the cosmological evolution of the universe from inflation to the final stage of reheating by the decay of the volume modulus. Inflation is driven by a blow-up mode and generating density perturbations of the correct size requires gaugino masses within the range $M \sim 10^4$-$10^5$ GeV. The crucial point is that if the K\"ahler metric of the matter fields has a suitable dependence on the inflaton, sleptons and squarks in split SUSY models can become tachyonic during inflation, while being non-tachyonic in the post-inflationary era. In consequence, the AD field can develop a large VEV during inflation and its subsequent motion can lead to the generation of a baryon asymmetry that survives the dilution due to entropy release in the final stage of reheating driven by the decay of the lightest modulus. The final reheating temperature is sufficiently high to allow thermal Higgsino-like DM with a mass around $1$ TeV.

\section*{Acknowledgements}

We would like to thank B. Dutta, G. Kane, A. Maharana, D. Marsh, F. Quevedo, K. Sinha and S. Watson for useful discussions. The work of R.A. is supported in part by NSF Grant PHY-1417510. This work was initiated at the Aspen Center for Physics, which is supported by NSF Grant PHY-1066293.

\end{document}